# HETEROGENEITY OF THE EDUCATIONAL SYSTEM: AN INTRODUCTION TO THE PROBLEM

*Authors*:
*Fuad Aleskerov*, HSE, School of Applied Mathematics and Information Science, Academic Supervisor
*Isak Froumin*, HSE, Institute of Education, Academic Supervisor
*Elena Kardanova*, HSE, Center of Education Quality Monitoring, Director
*Alina Ivanova*, HSE, Center of Education Quality Monitoring, junior researcher
*Alexey Myachin*, HSE, International Laboratory of Decision Choice and Analysis, researcher; Russian Academy of Sciences, Institute of Control Sciences, researcher
*Alexandra Serova*, HSE, Methodological Center, Supervisor
*Vyacheslav Yakuba*, Russian Academy of Sciences, Institute of Control Sciences, senior researcher

We analyze a heterogeneity of the educational system on the basis of one parameter: input grades of university students. We propose a mathematical model based on the construction of universities' interval order. We use the Hamming distance to evaluate the heterogeneity of the educational system, and the Unified State Examination (USE) scores of Russian students to illustrate the application of the model. We show that institutions taking weak students turn the whole system of universities into a poorly structured nonhomogeneous system. In contrast, after deleting the weakest part, the remaining set of universities becomes a well-structured system.

УДК 378:51
ББК 74.58в6

*Acknowledgments*. The results of the project "Differentiation of student trajectories in contemporary higher education (Studying the academic heterogeneity of the student contingent)", carried out within the framework of the Basic Research Program at the National Research University Higher School of Economics (HSE) in 2013, are also presented in this work.
F. Aleskerov and A. Myachin thank the International Laboratory DeCAn of HSE and Scientific Fund of HSE for partial financial support.

# Content

# 1. Introduction

One of the trends common in higher education around the world is its mass character. More and more freshmen with different social, economic and academic backgrounds enter universities every year. In turn, universities are increasingly competing with each other for the most talented students. Which universities attract the strongest student body? Which faculties attract the weakest students? To which extent has the system of higher education become non-homogeneous? And finally, how do universities differ in the quality and heterogeneity of their admissions? In this paper, we develop a methodology that helps us answer these questions.

# 2. Literature review

A survey of relevant literature reveals no special studies on the issue of system's heterogeneity in higher education. However, heterogeneity has been rather extensively studied as a phenomenon in the sphere of education. At least three possible viewpoints on the problem of heterogeneity can be derived from the existing studies: diversification of higher education institutions, selectivity of higher education institutions, and heterogeneity of the student population.

Institutional diversification of higher education institutions is considered as a response to the diversity of needs and requirements of society as a whole [Reichert, 2009]. The proponents of this view believe that the traditional institutions of higher education, such as classic universities, are no longer useful to meet the demands and interests of an increasingly heterogeneous student population, or to meet the requirements of a rapidly changing economy [Guri-Rosenblit, Sebkova, 2006]. Accordingly, the diversification of the educational system is the natural result of higher education becoming truly massified [Schofer, Meyer, 2005].

Critics of educational diversification argue that it does not satisfy the requirements of the entire heterogeneous population, especially of low-income students [Posselt et al., 2012]. The explanation is that diversity in higher education can turn into a polarized system with the competitive institutions on one end, and the rest of the institutions with low educational standards on the other end. So, while providing access to higher education for various groups of students, diversification of higher education possibly creates an additional hierarchy amongst institutions. This hierarchy is based on the prestige of different universities, and as a result it preserves an inequality that already exists in society today [Carnevale, Strohl, 2010].

The other phenomenon that reflects the heterogeneity problem is the selectivity of higher education institutions. Selectivity can be defined as a set of requirements aimed to separate entrants who can study at the university from those who cannot. In some countries educational systems can be characterized by the low selectivity of universities, for example, in France (except of Grandes Écoles). But despite the low entrance standards and quite heterogeneous set of students, selection of talented students and attrition of incapable students occurs, mainly during the learning process [Calmand et al., 2009]. If the university adheres to a rigid selection system, it will select the best candidates with the highest scores and the deepest academic backgrounds. At the same time, the strict selectivity and the prestige of some universities influences graduates' future opportunities, including income and professional status [Brewer et al., 1999; Datcher, Garman, 2005].

Nowadays, in some countries (e.g., in the USA) universities and colleges have to follow some requirements in order to contribute to the diversity of the student body. So, even prestigious colleges regularly use racial or legacy preferences, as well as preferences for talented athletes, in selecting their students [Pastine, Pastine, 2012]. Though high test scores are the most significant precondition for enrolment in the most selective universities, they do not guarantee a place in those institutions, and even the contingent of highly selective institutions can often appear academically non-uniform [Hurwitz, 2011].

Another possible viewpoint on the problem of heterogeneity is to consider the heterogeneity of the student body in terms of differences in the characteristics of students.

A number of studies have demonstrated that differences in family background have some effects on students' university choice and their educational achievements [McEwan, 2003; Woessmann, 2004; Brian, 2010]. Additionally, differences in peer socioeconomic status (SES) and their impact on student learning have been widely discussed [Evans et al., 1992; Robertson, Symons, 2003; van Ewijk, 2010]. A range of studies consider ethnicity and race as another possible source of heterogeneity, influencing student academic achievements [Baker et al., 2000; Braswell et al., 2001; Altschul et al., 2006].

Difference in academic ability is also a factor, setting students apart and affecting their performance. A number of researchers consider academic heterogeneity in terms of so-called peer-effects. They measure the effect that the average performance of one group of students has on another group of students [Sackerdote, 2001; Zimmerman, 2004; De Paola, Scoppa, 2010; Andrushak et al., 2012; Bielinska-Kwapisz, Brown, 2012]. Yet another important issue is how to estimate the impact of homogenous or heterogeneous group formation on students results [Lyle, 2008, Huang, 2009].

In all the studies discussed above, econometric tools are used. In particular different regression models (multiple, logistic, multilevel or quintile regression, etc.) try to estimate between students' heterogeneity and academic achievements. These models do not measure heterogeneity itself; researchers simply accept this phenomenon as a given. In order to measure the heterogeneity, they mainly refer to standard deviation, range or interquartile range, as well as the variance.

There have also been attempts to consider the problem of heterogeneity of the educational system on the country-level [Murdoch, 2002]. Using data from the CHEERS project (Careers after Higher Education: a European Research Survey) Murdoch assesses heterogeneity in selectivity between higher education institutions for each educational field within a set of countries (some European countries and Japan). To do so he takes the entry grade given by each graduate (on the scale of "high" (1), "medium" (2), and "low" (3)), then computes the mean for all the institutions and departments, and then finds the variance. Countries with a low variance are seen as homogeneous in terms of student selection across the different institutions, while countries with a high variance seem to be more heterogeneous in selectivity.

These studies do not consider heterogeneity in terms of the system as a whole. If analysis is performed on small samples (for example, within one student group), an interesting measure of heterogeneity is the Goodness of Heterogeneity Index [Graf, Bekele, 2006]. For each particular student a score is computed as the sum of all values of the student attributes (including interest in the subject, achievement motivation, level of performance in the subject, and others). The range of the index lies between 0 and a value that depends on the difference between the maximum and minimum sum of scores. The Goodness of Heterogeneity Index is equal to 0 if all students in a group have equal scores. A value less than 1 means unreasonable heterogeneity, i.e. student scores are at two extremes. And a value more than 1 indicates a reasonable level of heterogeneity. Under "reasonable level" authors suppose a combination of low, average and high student-scores. We doubt that this index can be used to evaluate heterogeneity in the problem under study.

In Fedriani and Moyano [2011] multidigraphs are used to evaluate the homogeneity/heterogeneity of schools. Authors develop the concept of heterogeneity in terms of the degree of hierarchy in these multidigraphs. Unfortunately, this approach cannot be used to analyze the system's heterogeneity, since the hierarchical structure is inherent to any system of education in the world.

It should be noted that some other tools from different fields of science could possibly be transferred to educational systems in order to measure their heterogeneity. For instance, in economics, attempts to formalize a systemic heterogeneity are mainly concerned to the evaluation of income inequality by the well-known Gini coefficient [Sudhir, Segal, 2008; Bosi, Seegmuller, 2006] or similar indices [Becker et al., 2005]. However, the use of Gini coefficient for our problem does not prove particularly useful, as the form of the corresponding Lorenz curve turns out to be linear.

In summary, while there are different approaches to consider and measure the problem of heterogeneity, almost all of them have some limitations and cannot be used to properly evaluate the heterogeneity of the educational system as a whole. Thus, we are introducing a new concept of estimating heterogeneity.

To do this we propose a mathematical model based on the construction of universities' interval order. Then we use the Unified State Examination (USE) scores of Russian students to illustrate how our measure of the system's heterogeneity works.

# 3. The model

## *3.1. The construction of interval order*

Denote the set of all universities as $A = \{1,...,n\}$. For each university $i \in A$ it is known the set of entrance grades $\{v_1^i,...,v_{k_i}^i\}$, where $k_i$ is the number of students in the university $i$.

First, we construct an interval order on the set $A$ of the universities under study. Assume that each university is defined by some interval $[a_i, b_i]$ depending on the values of the entrance exam grade. How these intervals are defined is shown below.

The interval order on A is constructed as
$$iPj \Leftrightarrow a_i > b_j \qquad (1)$$
In other words, if the left boundary for $i$ lies at the right of the right boundary of $j$, then we include the pair $(i,j)$ to the interval order.[1]

Let us consider an example how the interval order can be constructed for 5 universities $A = \{x_1,...,x_5\}$, and let the intervals for these universities be given as on Fig. 1.

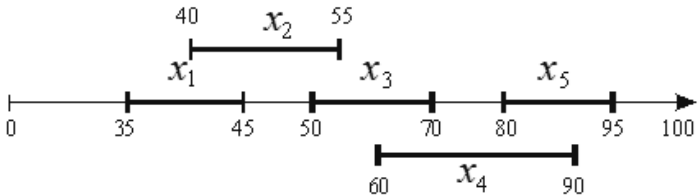

**Figure 1.** Intervals for 5 universities

The graph of the corresponding interval order constructed on $A$ with respect to the formula (1) is presented on Fig. 2.

---

[1]The notion of an interval order was introduced by Norbert Wiener [Wiener, 1914] (see also [Aleskerov et al., 1984], [Aleskerov et al., 2007]).

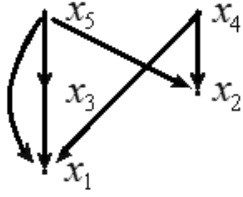

**Figure 2**. Graph of the interval order for the universities from Fig.1

The matrix $P$ (the incidence matrix) corresponding to this interval order is shown below. The matrix is constructed as follows: $P_{ij}=1$ $if$ $(i,j) \in P$, otherwise $P_{ij}=0$.

$$P_{id} = \begin{array}{c} \\ x_1 \\ x_2 \\ x_3 \\ x_4 \\ x_5 \end{array} \begin{pmatrix} x_1 & x_2 & x_3 & x_4 & x_5 \\ 0 & 0 & 0 & 0 & 0 \\ 0 & 0 & 0 & 0 & 0 \\ 1 & 0 & 0 & 0 & 0 \\ 1 & 1 & 0 & 0 & 0 \\ 1 & 1 & 1 & 0 & 0 \end{pmatrix}$$

To construct interval order for real universities several methods can be used because the intervals themselves can be constructed using different concepts.

For example, for the university we can construct the interval $[a_i, b_i]$ where $a_i$ is the minimal entrance exam grade, $a_i = \min_j v^i_j, j = 1,..,k_i$, and $b_i$ is the maximal entrance exam grade, $b_i = \max_j v^i_j, j = 1,..,k_i$, obtained by students of this very university.

The second version of the system of intervals for the university $i \in A$ is defined as $[\mu_i - \sigma_i, \mu_i + \sigma_i]$, where $\mu_i$ is the mean value of all entrance exam grades for this university $I$, $\mu_i = \frac{1}{k_i}\sum_j v^i_j$, and $\sigma_i$ is the standard deviation of the grades, $\sigma_i = \sqrt{\frac{1}{k_i}\sum_j (v^i_j - \mu_i)^2}$.

Following rather standard and widely used approach from mathematical statistics we apply the second method, i.e., the interval is defined as the mean value of grades ± standard deviation.

### 3.2. How to evaluate the heterogeneity of the educational system?

Suppose the interval order $P$ is constructed with respect to (1) using real data of the universities. Assume also that somehow the notion of ideal interval order $P_{id}$ is defined, which represents our vision on how the grades (students) are to be distributed among universities.

Then as the measure of heterogeneity the Hamming distance is used. This distance between two interval orders $B_1$ and $B_2$ is defined as

$$H(B_1, B_2) = \frac{1}{n(n-1)} \sum_{i,j} |b^1_{ij} - b^2_{ij}|, \qquad (2)$$

where $b^1_{ij}$ is equal to 1 if the pair $(i,j)$ belongs to the interval order $B_1$, otherwise $b^1_{ij}=0$; $b^2_{ij}$ is equal to 1 if the pair $(i,j)$ belongs to the interval order $B_2$, otherwise it is equal to 0.

As an example let us consider again 5 universities shown on Fig. 1. Assume that for these universities we construct being based on some model the ideal interval order $P_{id}$ shown on Fig. 3.

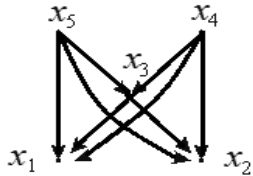

**Figure 3**. Graph for the ideal interval order

The corresponding incidence matrix for the ideal interval order is shown below

$$P_{id} = \begin{array}{c} \\ x_1 \\ x_2 \\ x_3 \\ x_4 \\ x_5 \end{array} \begin{array}{c} x_1 \ x_2 \ x_3 \ x_4 \ x_5 \\ \left( \begin{array}{ccccc} 0 & 0 & 0 & 0 & 0 \\ 0 & 0 & 0 & 0 & 0 \\ 1 & 1 & 0 & 0 & 0 \\ 1 & 1 & 1 & 0 & 0 \\ 1 & 1 & 1 & 0 & 0 \end{array} \right) \end{array}$$

Comparing the matrices for the real and the ideal interval orders and using (2) we can calculate the Hamming distance between two interval orders as $H(P, P_{id}) = 0.1$.

The Hamming distance characterized the deviance of the real system of universities from the ideal one: the fewer this distance is the better the real system of universities corresponds to our views about what it should be.

### *3.3. How to construct ideal interval orders?*

Below we will give several models how to construct ideal interval orders. Certainly they are not ideal in some mathematical sense, rather they correspond to our view what the ideal system is. The set of ideal universities can be constructed based on the different concepts that will be considered below.

We distinguish three main types of ideal systems: in the first two types we are oriented to some extent to the real data, while in the third type we introduce the concept of the desired system. This concept is based on our assumptions what the system of universities within specific areas of training should look like.

#### *3.3.1. Clustered system*
First, we suggest to construct an ideal interval order using some clustering of the universities based on their means of USE grades. So, we consider the mean value of grades for each university and cluster these values by *k*-means method for a given number of clusters. Different number of clusters can be used. We consider 3, 4 and 5 clusters. But only the 4 clusters case can be well interpreted for our case. Then for each university we define its ideal counterpart having the mean value $m^I$ as in the center of the cluster, and the interval for such ideal university is given as $[m^I - \sigma^I, m^I + \sigma^I]$, where $\sigma^I$ is the standard deviation for the corresponding cluster. Thus, each cluster contains several universities (equal to the number of universities placed in the cluster) with the same intervals.

Then we construct the interval order $P^I$ using the rule (1).

Let us consider a simple example of the real system of 4 universities with the parameters presented in the Table 1.

**Table 1.** Parameters of universities analyzed

| University | Mean USE score $\mu_i$ | Standard deviation $\sigma_i$ | Respective intervals | |
|---|---|---|---|---|
| | | | $\mu_I - \sigma_i$ | $\mu_I + \sigma_i$ |
| A | 60 | 5 | 55 | 65 |
| B | 65 | 5 | 60 | 70 |
| C | 80 | 3 | 77 | 83 |
| D | 90 | 5 | 85 | 95 |

The incidence matrix for the interval order for 'real' data is given in the Table 2 below.

**Table 2.** Incidence matrix for "real" data

| | A | B | C | D |
|---|---|---|---|---|
| A | 0 | 0 | 0 | 0 |
| B | 0 | 0 | 0 | 0 |
| C | 1 | 1 | 0 | 0 |
| D | 1 | 1 | 1 | 0 |

Let us construct $P^I$. For simplicity assume that there are only 2 clusters. One cluster contains A and B, and the other consists of C and D. The center of the first cluster is 62.5 with standard deviation of 3.54. The center of the second cluster is 85 with standard deviation of 7.07.

The intervals for this case and the incidence matrix for the interval order for clustered data are presented in the Tables 3 and 4.

**Table 3.** Intervals for clustered data

| Ideal points by clustering | Ideal point left | Ideal point right |
|---|---|---|
| iA | 58.96 | 66.04 |
| iB | 58.96 | 66.04 |
| iC | 77.93 | 92.07 |
| iD | 77.93 | 92.07 |

**Table 4**. The incidence matrix for the interval order for clustered data

|    | iA | iB | iC | iD |
|---|---|---|---|---|
| iA | 0 | 0 | 0 | 0 |
| iB | 0 | 0 | 0 | 0 |
| iC | 1 | 1 | 0 | 0 |
| iD | 1 | 1 | 0 | 0 |

Using formula (2) we can calculate the Hamming distance between real interval order and $P^I$ in this case. It is equal to 0.08.

### 3.3.2. Uniform system

The second method of constructing the ideal interval order deals with the centers of groups uniformly distributed over the set of universities. We divide into $k$ intervals of equal length the interval from the minimal value $m$ of real entrance grades, $m = \min_i m_i$, to the maximal value $M$ of these grades, $M = \max_i M_i$. Then we construct $k$ centers $c_i$ of these intervals (they will be on the equal distances from each other), and divide all universities into $k$ groups in accordance with the mean USE grades. The interval itself for each ideal group of universities is constructed as $[c_i - 0.001, c_i + 0.001]$. Note that these tiny intervals around such points are defined only to treat the data as an interval order and to avoid unnecessary complications with computer precision issues. Then we construct an interval order $P^{II}$ on the set of these ideal groups of universities using the rule (1).

As for a number $k$ of ideal groups of universities, after preliminary analysis our suggestion is to use $k = 4$. This number of groups can be easily interpreted: the group of very strong, elite universities admitting only high ability students and opposite, the group of weak universities with mainly low ability students; between these two polar groups there are two middle groups of universities. The difference between these middle groups is in ability level of students: students of one group are below the average, while students of another group are above the average.

Let us consider the example from the previous Section. Using the scores of Table 2 we construct four intervals [60.0,67.5), [67.5,75), [75,82.5), [82.5,90], to which the interval [60,90] is divided. So, the centers of these intervals are 63.75, 71.25, 78.75 and 86.25. Then the intervals for the ideal system are defined as [63.749, 63.751] [71.249, 71.251] [78.749, 78.751], [86.249, 86.251].

The intervals for this case and the incidence matrix for the interval order for uniformed data are presented in the Tables 5 and 6.

**Table 5.** Intervals for uniformed data

| University | Left bound | Right bound |
|---|---|---|
| uA | 63.749 | 63.751 |
| uB | 71.249 | 71.251 |
| uC | 78.749 | 78.751 |
| uD | 86.249 | 86.251 |

The incidence matrix for the interval order for this uniform system is given below.

**Table 6**. The incidence matrix for the interval order for clustered data

| University | uA | uB | uC | uD |
|---|---|---|---|---|
| uA | 0 | 0 | 0 | 0 |
| uB | 1 | 0 | 0 | 0 |
| uC | 1 | 1 | 0 | 0 |
| uD | 1 | 1 | 1 | 0 |

The Hamming distance between real interval order and uniform system is equal to 0.08.

*3.3.3. A desired system*

The third method of constructing the ideal interval order relates to our vision of the system of universities. The description of it will be given further separately for different groups of universities based on their majors. The ideal interval order $P^{III}$ can be constructed by the same way using the rule (1).

## 4. Hypothetical systems of universities

What is a system of universities that can be considered "desirable" for Russian higher education? What quality of applicants could ensure institutions of higher education for different educational needs? These issues became especially crucial when considering how the expansion of higher education affects all dimensions of society [Forrat, 2009; Belotserkovskii, 2011].

In this Section we try to construct a hypothetical model for the educational system based on an assumption about the stratification of institutions. We suppose that the educational system could have several levels of institutions based on the quality (and heterogeneity) of students they admit, i.e. from elite to ordinary universities. We presume that this structure, for example, best answers the current challenges of Russian society and the Russian economy.

First of all, we assume that the universities taking students with an average USE score of about 50 will not likely be able to train specialists prepared for the modern economy. Respectively, in our hypothetical system, within each integrated group of major there should not be institutions incapable of ensuring a proper quality of education. At the moment these lower-quality institutions function as social assistance; for example, they save certain young people from the military service or from unemployment. Throughout the last decades of mass higher education it has been considered an unwritten norm that in every region there is a university which "simply gives higher education" [Avanesov, 2013].

For our "hypothetical" systems, we would like to see universities of two main categories. Firstly, it could be a group of strong institutions that accept students with a very high USE. These institutions could, for example, focus on training high-class professionals with strategic vision, the future leaders of both secondary and tertiary level. These universities need to be engaged in scientific research and to provide breakthroughs in technologies of education. Secondly, it could also be a group of universities that provide mass, but high-quality, training of specialists, while taking into account regional specifics of the labor market. These may not necessarily be research universities. Yet these institutions can be focused on providing applicable competencies for their graduates.[2]

Let us consider in detail the hypothetical educational systems for technical, economic, medical and agricultural majors.

### *4.1. Hypothetical system of higher education institutions offering technical majors*

Originally, training in technical and engineering fields implies a certain level of preparation, mastery of a complex program including higher mathematics. It requires high scores on USE mathematics, which, in turn, becomes a threshold for applicants. In addition, the graduates of technical specialties have serious professional responsibilities, including public functions such as the prevention of potential man-made disasters. Therefore,

---

[2] In the European educational space the University of applied character, having only undergraduate programs may take high positions in the ranking. For example, Avans University of Applied Sciences (Netherlands) has been proclaimed the best major university of applied sciences in the national ranking at the end of 2012. <http://www.avans.nl/international/for-international-organisations/news/detail/2012/12/avans-once-again-the-best-in-the-netherlands>

for our model the lower bound of USE in the technical majors is set at the level of 60 points.[3]

A large segment of the universities (the universities of the second tier) should recruit students with scores in a relatively small corridor, from 60 to 70. This allows us to assume that the quality of education will be ensured not only by the quality of applicants at the entrance, but also due to the relatively small heterogeneity.

In this hypothetical system the first tier universities could also be a relatively small group, no more than 15–20 universities. The number of universities providing training within their group of majors is determined by the specific industry sectors, each of which needs 1–3 flagship universities. The group of first tier universities has to be focused on the preparation of high-level engineers. In our proposal, the first tier universities need students with average exam scores not lower than 70.[4]

Thus, in the group of technical majors we see a hypothetical situation that can schematically be presented as follows:

1) Group of leaders, no more than 10 universities, USE score of the admitted students should not be lower than 65–70 in the different majors.

2) Other higher education institutions (USE score of the admitted students should not be lower than 60).

## 4.2. *Hypothetical system of higher education institutions offering economic majors*

It has become commonplace to talk about an excess of economists and managers whom Russian universities send to the labor market, and who do not find work in their specialty. However, the labor market's demand for qualified specialists in economic, financial and managerial spheres is far from being satisfied. This apparent paradox signals a gap between the quality of training for this group of majors and the real expectations of employers regarding this quality. It follows that a significant optimization of the training system within Economics and Management is required. Currently the situation is as follows: almost 80% of universities, including technical and other non-core institutions, educate economists and managers. Even medical universities deem it necessary to have faculties of Economics and Management. As a result, the acceptance threshold for economic programs is not always high, nor is the quality of students' training [Egorshin et al., 2007].

---

[3] According to 2012 data within the integrated group of majors "Electronic Equipment, Radio Engineering and Communication" in general 64 higher education institutions took students with an average USE score below this level (authors' calculations).

[4] Our assumptions are not much far away from reality. For example, in the integrated group of majors "Electronic Engineering, Radio Engineering and Communications" there were only 5 such institutions according to our calculations based on the 2012 Monitoring data (see Section 5).

Therefore, the hypothetical system for economic training can include, from our point of view, three main sectors. First of all, we would allocate about 10% of the institutions to teach professional economists-analysts, with developed competences in strategic planning and management. These universities should take a contingent of standout students with an average USE score not lower than 75. These could be higher education institutions which prepare high-class analysts to work in federal and global labor markets.

At the other end of the system we can assume that about 20% of the institutions will prepare good and strong applied specialists, combining practical skills (of the former graduates of the secondary vocational education) with more developed competences, including goal-setting, professional responsibility, open-mindedness, and with fundamental training in basic disciplines. Applied baccalaureate programs can offer such results, closing the niche of the secondary vocational education, which has been eroded by the accessibility of mass higher education. The individual average score of Unified State Examination of the students accepted to the economic programs in these universities should not be lower than 60.

Finally, between these two poles, sits the largest group of higher education institutions – about 70% of all universities within the system. These universities have to be focused on preparation of specialists for regional labor markets, for various sectors of the economy, and for industry. Accordingly, students who come to these universities are also located between the two poles mentioned above. Their average level of training, expressed in USE mean score, is expected to be in the range from 65 to 74.

Thus, the hypothetical system of higher education institutions conducting preparation of economists can be presented as follows:

1) A group of elite universities that train managers, strategists, high-class analysts (about 10% of universities). Scores of admitted students should not fall below 75.

2) A group of strong universities that train strong professionals for regional labor markets (about 70% of universities). Scores of admitted students from 65 to 74.

3) Group of acceptable quality universities preparing bachelors in applied programs (about 20% of universities). Scores of admitted students should not be lower than 60.

## 4.3. Hypothetical system of higher education institutions offering agrarian majors

Young people in Russia, unfortunately, do not see prospects for agricultural education in the current day. Agricultural specialties are no longer in demand, even in specialized institutions. Often the choice of entrants, if it is not due to family tradition [Filonenko, Lepin, 2013] or pronounced vocation, is far from their priority. Students choose agricultural specialties when it is impossible to get an education in other, more preferable specialties. This is partly due to a decrease in demand for specialists in this field (with the arrival of technical progress to agriculture), and the general trend of migration toward cities. Thus, it is not possible to provide a high starting level of applicants. Yet, it is clear that admission of applicants with USE score below 50 effectively means a refusal to provide any quality of training.

Therefore, in this hypothetical system, we see opportunities for a large group of universities oriented toward students with scores from 50 to 60. These students may well receive an acceptable level of training and become experienced agriculture workers needed in the regions.

Another relatively small group of universities should provide admission to students with USE scores not lower than 60. These universities can hardly be called leading (though it does not exclude the presence of several universities in this group – the flagships with great research potential). But they can be focused on training managers, strategic planners, and a small number of researchers for the agricultural industry.

Thus, the hypothetical situation for a group of agricultural specialties is as follows:

1) A group of strong universities that prepare agriculture managers at various levels (about 10% of universities). USE scores of applicants being at least 60.

2) The remaining universities. USE scores of applicants not lower than 50.

## *4.4. Hypothetical system of higher education institutions offering medical majors*

Traditionally, medical education is highly-demanded and well-recognized in society (sometimes despite small salaries and severe working conditions in this sphere). Medical universities almost never have problems with competition among students, and traditionally select the best entrants with high average scores. In fact, the hypothetical model for healthcare can be structurally close to the existing real situation: we would distinguish three groups of universities, and the difference between applicants' input quality would be significant among them.[5] Groups may differ more by external factors than by the quality of applicants.

First of all, we would have identified a group of universities that are concentrated in the lower boundary, with student scores from 65 to 75. We assume that these are regional universities. They could deal with those students who did not dare to apply for medical specialities at more prestigious universities. In terms of the hypothetical system, they could cover the needs of the regional labour markets in health personnel.

At the other extreme we could put a small group of leaders, who have all the opportunities and resources for science and technology development both in education and in the professional sphere. They can select the best of the best. They can provide high quality education, due to a homogeneous and well prepared contingent, and the actual quality of the educational process. The graduates of such universities are in high demand on the federal labor market.

And third, there can be a large intermediate group of universities. This group is likely to include medical faculties of big multidisciplinary universities, unless they are based on older specialized institutions. This may also be mostly regional universities, which will provide the highly-qualified medical personnel, but for the local, not federal labor market.

---

[5] The variation is about 5 points in either direction from the student individual average score.

The hypothetical situation for the group of medical specialties would be as follows:

1) A group of leading universities (about 10% of universities). Students' scores are not lower than 75.

2) A group of strong universities. Students score is at least 70.

3) Good universities. Students' scores are not lower than 65.

To conclude, we realize that such a stratified structure may have implications regarding the perpetuation of societal inequality. Nevertheless, we would like to emphasize that this is only a hypothetical vision, a modeled situation with two main goals: to show the deviance of the real educational system from this model, and to demonstrate how our method works.

## 5. The data

The empirical data for this research was obtained from the monitoring database "Quality of students' enrollment" which has been carrying out by the National Research University "Higher School of Economics" (NRU HSE)[6] and Russian Information Agency (RIA News).

The database contains mean scores of Unified State Examination (USE) of full-time students enrolled in 2011–2013 on the Bachelor Degree Programs, as well as the information on the forms of admission (on competition, out of competition, on a tuition-based basis or on a state-financed basis, etc.). We use the data for 2012.

We examine the differences in the heterogeneity in the context of integrated groups of majors. As a basis for the allocation of such groups the official list of majors in higher education is used.[7]

This document contains 27 integrated groups of majors (Bachelor programs). We examine four of them: Economy and Management (as the most popular), Electronic Engineering, Radio Engineering and Communications (as an example of Engineering specialties), as well as Agriculture and Fisheries

---

[6] Website of the Monitoring "Quality of Universities Admissions – 2012". <http://www.hse.ru/ege/second_section2012/>

[7] Annex 1 to the Order of the Ministry of Education and Science of the Russian Federation of February 17, 2011 No. 201. <http://base.garant.ru/55171451/>

and Medicine (polar groups of majors, in terms of the quality of students they attract).

All the Monitoring data that we use in this study are open data. The database for Monitoring consists, in general, of the average scores of students admitted to the 1st year bachelor programs, full-time form of training.[8]

Unfortunately, not all universities have posted on their sites full information about admission. At a result, in our database, there are a small number of missing values. How the missing values are filled in see in Appendix 1.

It should be also noted, that for purposes of this research we do not use the whole set of the data we have. In the Table 7 the sets of initial and used data are described. Removal from the general sample occurs because of a small number of students or big percentage of missing data (for details see criteria 7 and 8 of Appendix 1).

**Table 7.** The general characteristics of the data used

| Group of Majors | Electronic Engineering | | Economy and Management | | Agriculture and Fishery | | Healthcare | |
|---|---|---|---|---|---|---|---|---|
| Data | Initial data | Used data | Initial data | Used data | Initial data | Used data | Initial data | Used data |
| Total number of universities | 109 | 105 | 408 | 379 | 94 | 85 | 77 | 77 |
| Total number of enrolled students | 10199 | 9759 | 93576 | 91336 | 18477 | 17889 | 36686 | 36686 |

Further we present the data of universities sample within each integrated group of majors, that have not been deleted from the general sample.

The main descriptive characteristics of universities in each of our four data sets are presented in Appendix 2, while Table 8 provides the generalized information of universities and groups of majors under study. It should be pointed that Table 8 shows the data corresponding to the criteria 7 and 8 in Appendix 1. The agricultural universities are exceptions, where criterion 8 is shifted: minimum number of students in this category is 8 (instead of 15).

---

[8] See <http://www.hse.ru/ege/second_section2012/meth>.

**Table 8.** An overview of the data

| Group of Majors | Electronic Engineering | Economy and Management | Agriculture and Fishery | Healthcare |
|---|---|---|---|---|
| Total number of universities | 105 | 379 | 85 | 77 |
| Total number of enrolled students | 9759 | 91336 | 17889 | 36686 |
| Score mean (range) | 47,39–76,97 | 46,32–83,97 | 39,77–64,45 | 56,61–89,32 |
| Standard deviation (range) | 2,9–17 | 3,07–17, 23 | 3,09–13,16 | 6,38–14,3 |
| Score median (range) | 46,3–80 | 43,7–93,3 | 39,7–64,7 | 55–93 |

Further, based on the proposed models the heterogeneity of universities within each of the selected groups of majors will be investigated.

## 6. Results and discussion

### 6.1. Universities within the integrated group of majors "Electronic Engineering, Radio Engineering and Communications"

We will consider the heterogeneity of the system of universities that offer courses in the integrated group of majors called here "Electronic Engineering". 105 universities from this field were studied. We did not divide the students in groups of tuition-based and state-financed students because only a small amount of tuition-based students study within this group of majors (see Fig. A1 in Appendix 2).

For a real system of universities, an interval order was built where the interval for each university with the average USE scores of enrolled students ± standard deviation of these scores were calculated.

Then, according to the description in Sections 3 and 4, three types of ideal systems are constructed.[9]

---

[9] In addition, we have reviewed other systems. For example, ideally abstract situations where universities were distributed uniformly with small intervals (with a diameter of not more than 0.001), and when the universities were located in real points with small intervals (with a diameter of not more than 0.001). But the detailed analysis revealed that these systems are less convenient for an interpretation and do not fit the available data.

1) Clustered system (4 clusters, subsection 3.3.1).
2) Uniform system (4 groups, subsection 3.3.2).
3) Hypothetical system (subsection 3.3.3).

When constructing the ideal system using the first method, a clustering of universities based on USE scores and using k-means for a given number of clusters (4) is made. The results are shown in Table 9.

**Table 9.** Clustered system (universities within the group of majors "Electronic Engineering")

| Cluster ID | Mean | St. Dev. | Count |
|---|---|---|---|
| 4 | 58.43 | 0.89 | 11 |
| 3 | 60.38 | 0.46 | 30 |
| 2 | 61.73 | 0.38 | 45 |
| 1 | 63.60 | 0.72 | 19 |

Unfortunately, the clustering of the system does not allow us to allocate well interpreted clusters of higher education institutions. In general, students in this group of majors are rather homogeneous compared to other integrated groups. The level of difficulty in the technical specialties initially assumes a certain degree of preparation in mathematics and physics, which determines the requirements for examinations in these subjects. In summary, this group of university freshmen consists mostly of "B" students.

Then each university is replaced by its ideal counterpart – the center of this cluster. For each university the interval is also constructed as the center (mean value) of the cluster ± standard deviation for the corresponding cluster, and further the interval order based on formula (1) will be defined.

Comparing the matrices $P$ for real and ideal interval orders and using formula (2) we can calculate the Hamming distance between two interval orders, $H(P, P_{id}) = 0.35$.

When constructing an ideal system by the second method, we first made a partition of all universities in 4 groups. The interval from the minimal value $m = 47.39$ of entrance grades to the maximal value $M = 76.97$ of these grades, we divided into 4 intervals of equal length. Each university was assigned to one of the 4 groups, according to its USE score. Characteristics of the uniform division into 4 classes of the universities system within the integrated group of majors Electronic Engineering are presented in the Table 10.

**Table 10.** Uniform system (universities within the group of majors "Electronic Engineering")

| Number | Left | Right | Center | Mean | St. Dev. | Count |
|---|---|---|---|---|---|---|
| 1 | 47.39 | 54.78 | 51.08 | 51.81 | 3.31 | 27 |
| 2 | 54.78 | 62.18 | 58.48 | 58.16 | 1.66 | 47 |
| 3 | 62.18 | 69.57 | 65.87 | 64.65 | 1.91 | 25 |
| 4 | 69.57 | 76.97 | 73.27 | 72.45 | 2.23 | 6 |

Further, each institution from the relevant group is replaced with its ideal counterpart $c_i$ – the center of the interval of the corresponding group. The interval itself for each ideal group of universities is constructed as $[c_i - 0.001, c_i + 0.001]$. Then we construct an interval order on the set of these ideal groups of universities using the rule (1).

Comparing the matrices $P$ for real and ideal interval orders and using formula (2) we can calculate the Hamming distance between two interval orders, $H(P, P_{id}) = 0.30$.

Finally, when constructing an ideal system by the third method we start from the hypothetical system of universities, admitting students in technical specialties, described in subsection 3.3.3, which can be schematically represented as follows:

1) Group of leaders, no more than 10 universities (an average USE score of the admitted students should not be lower than 65–70 in the different majors).

2) Other higher education institutions (an average USE score of the admitted students should not be lower than 60).

To explore the possibility of approximation to such a system, we have divided the entire system of universities into 3 groups – the most powerful, elite (scores above 70), the weakest (scores below 55, i.e. "C" students) and others (scores are from 55 to 70). Characteristics of the splitting are presented in Table 11.

**Table 11.** The 3 groups of universities (universities within the group of majors "Electronic Engineering")

| The scores' interval | Mean | St. Dev. | Count |
|---|---|---|---|
| >70 | 72.99 | 3.40 | 5 |
| [55;70] | 59.72 | 3.97 | 80 |
| <55 | 50.89 | 1.83 | 20 |

We will consider this system as a prototype for the ideal system. Each institution of the relevant group we replace by the ideal counterpart – the mean value of USE scores for this group. For each group we construct an interval order based on the mean value of scores in the group ± standard deviation for the corresponding group.

Comparing the matrices *P* for real and ideal interval orders and using formula (2) we can calculate the Hamming distance between two interval orders: $H(P, P_{id}) = 0.15$.

Comparing the values of the Hamming distance for all three types of ideal systems, we conclude that the latter system which is based on a splitting of all universities into 3 groups – elite, weak and others in the best way reflects the real situation for the group of majors Electronic Engineering, Radio Engineering and Communications.

However, as noted earlier, the recommended lower limit of the average USE scores for technical training areas lies at the level of 60 points. The smaller score cannot guarantee a sufficient level of initial preparation for learning on engineering programs. And in our system, there is a group of 20 universities, where the average score of enrolled students is below 55. Deleting this group of universities from the analysis gives a very interesting result: the Hamming distance between real and ideal interval orders becomes $H(P, P_{id}) = 0.04$.

Thus, if we exclude from the system very weak universities, the remaining pool of universities is becoming a well-coordinated system, i.e., our system of universities within the group of majors Electronic Engineering, Radio Engineering and Communications comes close to the desired system.

## 6.2. Universities within the integrated group of majors "Economics and Management"

Let us consider the heterogeneity of the system of universities within the integrated group of majors "Economics and Management". In this group 379 universities are studied, 342 of them conduct the admission of tuition-based students.

For a real system of universities the interval order was built, where the interval for each university with an average USE scores of enrolled students ± standard deviation of these scores were calculated. Then for each case three types of ideal systems were build (similarly to the way it was done for a group of technical universities (Section 6.1)): Clustered system, Uniform system and Desired system.

First, we examine all the students entirely and then separately for state-funded and tuition-based forms of training.

When constructing the ideal system by the first way the clustering of 379 universities based on the average USE scores and using $k$-means for a given number of four clusters is made. The results are shown in Table 12.

**Table 12.** Clustered system ("Economics and Management", state-funded + tuition-based)

| Number | Cluster ID | Mean  | St. Dev. | Count |
|--------|------------|-------|----------|-------|
| 1      | 4          | 51.49 | 2.02     | 86    |
| 2      | 3          | 57.20 | 1.64     | 169   |
| 3      | 1          | 63.71 | 2.41     | 92    |
| 4      | 2          | 74.20 | 4.08     | 32    |

The clustering of institutions from this system has allowed us to allocate various groups of universities accurately. The first group is formed by institutions taking weak students, generally mediocre pupils. We have 86 such universities, i.e. nearly a quarter. The second group consists of universities, taking overall "B" students, but relatively weak ones. This group is the largest, it has 169 universities. A quarter of universities (92) form a group, taking "B" – students, but quite strong in average. Finally, the last group consists of universities, attracting mainly standouts (both on state-funded, and on fee-paying form of training). There are 32 such universities.

In general the contingent coming on the economic departments of universities is characterized by the highest degree of heterogeneity (from the discussed groups of majors). This is partly due to the difference in scores of fee-paying and state-financed students. But to a large extent, this heterogeneity may occur due to the huge amount of non-core universities which have economic departments and do not always hold the high bar of admission.

Then each university is replaced by its ideal counterpart – the center of this cluster. For each cluster the interval order is constructed, as the center (mean value) of the cluster ± standard deviation for the corresponding cluster.

Comparing the matrices $P$ for real and ideal interval orders and using formula (2) we can calculate the Hamming distance between two interval orders: $H(P, P_{id}) = 0.31$.

Next, we performed a similar analysis, but separately for state-funded form of training and for tuition-based form. Tables 13 and 14 present the results of the clustering based on the average score of enrolled students, respectively, for the state-financed form (Table 13) and fee-paying forms of education (Table 14).

**Table 13.** Clustered system ("Economics and Management", state-funded)

| Number | Cluster ID | Mean | St.Dev. | Count |
|---|---|---|---|---|
| 1 | 4 | 57.38 | 2.99 | 40 |
| 2 | 3 | 64.48 | 1.69 | 117 |
| 3 | 2 | 70.52 | 1.85 | 126 |
| 4 | 1 | 79.00 | 3.95 | 45 |

**Table 14.** Clustered system ("Economics and Management", tuition-based)

| Number | Cluster ID | Mean | St.Dev. | Count |
|---|---|---|---|---|
| 1 | 4 | 49.85 | 1.81 | 100 |
| 2 | 1 | 54.32 | 1.22 | 157 |
| 3 | 2 | 58.83 | 1.91 | 74 |
| 4 | 3 | 69.16 | 3.98 | 11 |

In general, the state-funded form of training consists of more strong students in comparison to tuition-based form. Yet for each form this cluster solution looks quite convincing. So, for the state-funded form of training we would allocate two groups of higher education institutions attracting very strong students (groups 3 and 4 in Table 13). Together, it is almost half of all universities within this group of majors. A group of universities that accept state-financed students with their average scores close to "C"-students (group 1 in Table 13) is quite small, about 10% of institutions. For the fee-paying form of education the situation is opposite. The absolute majority of higher education institutions, 75% of universities, (groups 1 and 2 in Table 14) accept very weak students. Only 11 universities (3% of universities having fee-paying students) are able to attract on paid programs relatively strong students (group 4 in Table 14).

Next, in the same manner as described previously, the Hamming distance is evaluated. The result was that for both tuition-based and state-financed forms of admission the Hamming distance between two interval orders is $H(P, P_{id}) = 0.31$. Thus, the Hamming distance does not depend on the form of admission: despite various cluster solutions for state-financed and fee-paying forms, the real system's deviation from the ideal one is the same. Further we will consider all set of students entirely, without carrying out division into the state-financed and tuition-based forms of education.

When constructing an ideal system by the second method, we first made a partition of all universities in 4 groups. Characteristics of the uniform partition of this universities system into four classes are presented in Table 15.

**Table 15.** Uniform system ("Economics and Management", state-funded+ tuition-based)

| Number | Left | Right | Center | Mean | St. Dev. | Count |
|---|---|---|---|---|---|---|
| 1 | 46.3 | 55.7 | 51.03 | 52.71 | 2.42 | 128 |
| 2 | 55.7 | 65.1 | 60.44 | 59.34 | 2.50 | 190 |
| 3 | 65.1 | 74.6 | 69.85 | 68.93 | 2.87 | 51 |
| 4 | 74.6 | 84.0 | 79.26 | 79.38 | 2.82 | 10 |

We may notice that only 10 institutions take mainly strong students (with the average USE score higher than 74.6), while 128 institutions take weak, mainly mediocre students (with an average score lower than 55.7).

Further, each institution from the relevant group is replaced with its "ideal" counterpart $c_i$ – the center of the interval of the corresponding group. The interval itself for each ideal group of universities is constructed as $[c_i - 0.001, c_i + 0.001]$. Then we construct an interval order on the set of these ideal groups of universities using the rule (1).

Comparing the matrices $P$ for real and ideal interval orders and using formula (2) we can calculate the Hamming distance between two interval orders: $H(P, P_{id}) = 0.28$.

Finally, when constructing an ideal system by the third method we start from the hypothetical system of universities, admitting students in economic specialties, as described in subsection 3.2, which can be schematically represented as follows:

1) A group of elite universities that train managers, strategists, high-class analysts (about 10% of universities, the average score of the whole contingent of enrolled students should not fall below 75).

2) A group of strong universities that train strong professionals for regional labor markets (about 70% of universities, the average score from 65 to 74).

3) Group of satisfactory universities preparing bachelors on applied programs (about 20% of universities, an average USE score of the admitted contingent should not be lower than 60).

To explore the possibility of approximation to ideal system, we divide the entire system of universities into four groups – the most powerful, elite (average score above 75), strong universities (average score from 65 to 75), just good institutions (from 55 to 65) and the rest (the average score is below 55). Characteristics of the splitting are presented in Table 16.

We will consider this system as a prototype for ideal system. Each institution of the relevant group we replace by the "ideal" counterpart – the average value of USE scores for this group. For each group, we construct an interval order based on the mean value of scores in group ± standard deviation for the corresponding group.

**Table 16.** The splitting into 4 groups (universities within the group of majors "Economics and Management")

| The scores' interval | Mean | St. Dev | Number of universities |
|---|---|---|---|
| >75 | 79.38 | 2.82 | 10 (3%) |
| (65;75] | 68.85 | 3.29 | 52 (14%) |
| (55;65] | 58.76 | 2.67 | 220 (58%) |
| <=55 | 51.85 | 1.26 | 97 (25%) |

Comparing the matrices $P$ for real and ideal interval orders and using formula (2) we can calculate the Hamming distance between two interval orders: $H(P, P_{id}) = 0.26$.

Comparing the values of the Hamming distance for all three types of ideal systems, we conclude that the latter system which is based on a splitting of all universities into four groups (elite, strong, ordinary and weak universities) in the best way reflects the real situation for the group of majors "Economics and management".

Of course, the number of strong institutions in this group is much smaller than we would like – just 17% versus desired 80%. It is therefore necessary to improve the quality of admission for all universities in this group of majors. However, in Table 16, there is a group of 97 universities, gaining mostly very weak students (average USE score below 55). As noted earlier, in this group should not be universities taking students with an average score below 60. Similarly to the way it was done for a group of technical universities, we will "delete" from our system a group of weak universities. Deleting this group of universities from the analysis gives a very interesting result: the Hamming distance between real and ideal interval orders becomes $H(P, P_{id}) = 0.16$.

Thus, if we exclude from the system very weak universities, the remaining set of universities is becoming a well-structured system, however, still far from the desirable one.

### 6.3. Universities within the integrated group of majors "Agriculture and Fishery"

We will consider the heterogeneity of the system of universities within the integrated group of majors "Agriculture and Fishery". 85 universities from this field were studied. We did not divide the student contingent in 2 groups of fee-paying and state-financed students. Our decision can be explained by small amount of fee-paying students within this group of majors. According to our database only 7% of universities have fee-paying admission, and 60% of them admit less than 10 students (see Fig. A9 in Appendix 2).

For a real system of universities the interval order was built, where the interval for each university with an average USE scores of enrolled students ± standard deviation of these scores were calculated. Then for each case three types of ideal systems were built: Clustered system, Uniform system and Desired system.

When we construct the ideal system by the first way, we cluster universities based on the average USE scores and use the method of *k*-means for a given number of four clusters. The results are shown in Table 17.

**Table 17.** Clustered system (universities within group of majors "Agriculture")

| Number | Cluster ID | Mean | St. Dev. | Count |
|---|---|---|---|---|
| 1 | 3 | 43.46 | 1.96 | 11 |
| 2 | 1 | 49.80 | 1.51 | 40 |
| 3 | 2 | 54.31 | 1.32 | 28 |
| 4 | 4 | 60.77 | 2.02 | 6 |

All universities conducting enrollment of students to agriculture programs generally attract quite uniform, but the weak contingent of students. There are no universities in this group, which attracts in average students-standouts. Unfortunately, the students enrolled in agricultural majors for the most part cannot cross the border of 55 USE scores. Thus 51 universities (60%) are gaining very weak students (groups 1 and 2 in Table 17).

Then each university from the corresponding cluster is replaced by its ideal counterpart – the center of this cluster. For each university also the interval is constructed, as the center (mean value) of the cluster ± standard deviation for the corresponding cluster. And further the interval order based on formula (1) will be defined.

Comparing the matrices $P$ for real and ideal interval orders and using formula (2) we can calculate the Hamming distance between two interval orders: $H(P, P_{id}) = 0.31$.

When constructing an ideal system by the second method, we first made a partition of all universities into four groups. Characteristics of uniform partition of this universities' system are presented in Table 18.

**Table 18.** Uniform system (universities within the group of majors "Agriculture")

| Number | Left | Right | Center | Mean | St. Dev. | Count |
|---|---|---|---|---|---|---|
| 1 | 39.8 | 45.9 | 42.86 | 43.20 | 1.86 | 10 |
| 2 | 45.9 | 52.1 | 49.03 | 49.71 | 1.60 | 41 |
| 3 | 52.1 | 58.3 | 55.19 | 54.31 | 1.32 | 28 |
| 4 | 58.3 | 64.5 | 61.36 | 60.77 | 2.02 | 6 |

It is interesting to note that for this group of majors the uniform partition practically coincides with the cluster solution, which can be explained by the low variation in scores of enrolled students.

Further, each institution from the relevant group is replaced with its "ideal" counterpart $c_i$ – the center of the interval of the corresponding group. The interval itself for each ideal group of universities is constructed as $[c_i - 0.001, c_i + 0.001]$. Then we construct an interval order on the set of these ideal groups of universities using the rule (1).

Comparing the matrices $P$ for real and ideal interval orders and using formula (2) we can calculate the Hamming distance between two interval orders, $H(P, P_{id}) = 0.31$.

Finally, when constructing an ideal system by the third method we are based on the hypothetical system of universities, admitting students on agriculture specialties, as described in subsection 3.3, which can be schematically represented as follows

1) A group of strong universities that prepare agriculture managers at various levels (about 10% of universities; the average USE score of student contingent at least 60).

2) The remaining universities (an average USE score of students not lower than 50).

To study the possibility of approximation to such a system, we split the entire system of universities into 3 groups – good-quality universities (average score above 60), the weakest (average score below 50, i.e. "C" students) and others (average score of 50 to 60). Characteristics of the splitting are presented in the Table 19.

**Table 19.** The splitting into 3 groups (universities within the group of majors "Agriculture")

| The scores' interval | Mean  | St. Dev | Count    |
|----------------------|-------|---------|----------|
| >60                  | 61.69 | 1.84    | 4 (5%)   |
| [50;60]              | 53.28 | 2.2     | 49 (58%) |
| <50                  | 46.82 | 2.81    | 32 (43%) |

We will consider this system as a prototype for ideal system. Each university of the relevant group we replace by its "ideal" counterpart – the average value of USE scores for this group. Then for each group we construct an interval order as the mean value of scores in the group ± standard deviation for the corresponding group.

Comparing the matrices $P$ for real and ideal interval orders and using formula (2) we can calculate the Hamming distance between two interval orders, $H(P, P_{id}) = 0.25$

Comparing the values of the Hamming distance for all three types of ideal systems, we conclude that the latter system which is based on the splitting of all universities into 3 groups is the best. It reflects in the best way the real situation for the group of majors "Agriculture and Fishery".

However, as we have noted earlier, the recommended lower limit of the average USE scores for agriculture majors lies at the level of 50 points. The smaller score cannot guarantee a sufficient level of initial preparation to receive good-quality education. And in our system, there is a group of 32 universities, where the average score of enrolled students is below 50. Deleting this group of universities from the analysis gives a very interesting result: the Hamming distance between real and ideal interval orders becomes $H(P, P_{id}) = 0.07$.

Thus, if we exclude from the system very weak universities, the remaining pool of universities turns to a better coordinated system. It is necessary to tighten weak students to the minimum acceptable level, at least, not lower than 50 points. Possibly it could be done through the system of additional education.

### 6.4. Universities within the integrated group of majors "Healthcare"

Now we consider the heterogeneity of the system of universities within the integrated group of majors "Healthcare". 77 universities from this field were studied. We did not divide the student contingent in two groups of fee-paying and state-financed students, because there are not a lot of fee-paying students within this group of majors (See Appendix 2, Fig. A13).

For a real system of universities the interval order was built, where the interval for each university with an average USE scores of enrolled students ± standard deviation of these scores were calculated. Then for each case three types of ideal systems were build (as well as for other groups of majors): Clustered system, Uniform system and Desired system.

Then we cluster universities based on the average USE scores and use the method of *k*-means for a given number of clusters. The results are shown in Table 20.

**Table 20.** Clustered system (universities within group of majors "Healthcare")

| Number | Cluster ID | Mean  | St. Dev. | Count |
|--------|-----------|-------|----------|-------|
| 1      | 4         | 61.01 | 1.99     | 13    |
| 2      | 3         | 66.18 | 1.27     | 22    |
| 3      | 2         | 71.13 | 2.07     | 31    |
| 4      | 1         | 79.90 | 3.65     | 11    |

Students choosing medical specialties look as very strong and homogeneous cohort. So, here we see the division of universities rather on two groups and two subgroups in each of them. Essentially universities within this field could be divided on "A" students (excellent students) and "B" students (good students). At the same time it is possible to allocate in each group the subgroups of average performers and universities-leaders.

Then each university from the corresponding cluster is replaced by its ideal counterpart – the center of this cluster. For each university also the interval is constructed, as the center (mean value) of the cluster ± standard deviation for the corresponding cluster. And further the interval order based on formula (1) will be defined.

Comparing the matrices $P$ for real and ideal interval orders and using formula (2) we can calculate the Hamming distance between two interval orders, $H(P, P_{id}) = 0.34$.

When constructing an ideal system by the second method, we first made a partition of all universities in four groups. Characteristics of uniform partition of this university system are presented in Table 21.

**Table 21.** Uniform system (universities within the integrated group of majors "Healthcare")

| Number | Left | Right | Center | Mean | St. Dev. | Count |
|---|---|---|---|---|---|---|
| 1 | 56.6 | 64.8 | 60.7 | 61.8 | 2.3 | 17 |
| 2 | 64.8 | 73.0 | 68.9 | 68.8 | 2.2 | 43 |
| 3 | 73.0 | 81.1 | 77.1 | 76.9 | 2.4 | 15 |
| 4 | 81.1 | 89.3 | 85.2 | 85.9 | 4.8 | 2 |

Further, each institution from the relevant group is replaced with its "ideal" counterpart $c_i$ – the center of the interval of the corresponding group. The interval itself for each ideal group of universities is constructed as $[c_i - 0.001, c_i + 0.001]$. Then we construct an interval order on the set of these ideal groups of universities using the rule (1).

Comparing the matrices $P$ for real and ideal interval orders and using formula (2) we can calculate the Hamming distance between two interval orders $H(P, P_{id}) = 0.29$.

In the end we construct an ideal system by the third method. We also start from the hypothetical system of universities, taking students on medical specialties, as described in subsection 3.4. Schematically it could be done as follows

1) A group of leading universities (about 10% of universities, students' score are not lower than 75).
2) A group of strong universities (students' score is at least 70).
3) Simply good universities (students' scores are not lower then 65).

In order to study the possibility to be closer to such a system, we split the entire system of universities into four groups – leading, elite institutions (average score higher than 75), good-quality universities (average score of enrolled students varies from 65 to 75), ordinary universities (average score of enrolled students varies from 60 to 65) and others (average score not lower than 60). The results of this splitting are presented in Table 22.

**Table 22.** The splitting into four classes (universities within the integrated group of majors "Healthcare")

| The scores' interval | Mean | St. Dev. | Count |
|---|---|---|---|
| >=75 | 79.16 | 3.79 | 13 |
| (65;75] | 69.41 | 2.65 | 45 |
| [60;65] | 63.24 | 1.44 | 14 |
| <60 | 59.03 | 1.38 | 5 |

We will consider this system as a prototype for ideal system. Each university of the relevant group we replace by its "ideal" counterpart – the average value of USE scores for this group. Then for each group we construct an interval order as the mean value of scores in the group ± standard deviation for the corresponding group.

Comparing the matrices $P$ for real and ideal interval orders and using formula (2) we can calculate the Hamming distance between two interval orders, $H(P, P_{id}) = 0.29$.

And again, if we compare the values of the Hamming distance for all three types of ideal systems, we can conclude that the latter system which is based on a splitting of all universities into 4 groups is the best. It reflects in the best way the real situation for the group of majors "Healthcare".

But as we have noted earlier, the desirable lower limit of the average USE scores for medical majors lies at the quite high level of 60 points. The smaller score seems cannot guarantee a sufficient level of initial preparation for training the future health workers. The current system of universities has 5 universities, which admit students with the average score below 60. Deleting this group of universities from the analysis gives though less noticeable than in other systems, but also an important result: the Hamming distance between real and ideal interval orders becomes $H(P, P_{id}) = 0.26$.

Thus, if we exclude from the system institutions with weakest students, the remaining set of universities becomes well-structured system as in other cases. Certainly on the example of integrated group of majors "Healthcare" we do not see any essential improvement of the system after excluding weak institutions. We can suppose here two possible explanations. Firstly, this is because the real and the ideal systems do not have principal differences. And secondly, the number of universities which we remove from the system is the smallest compared to other groups of majors.

## 6.5. Discussion

We compared the real and "ideal" systems of universities for all of the integrated groups of majors examined in our study. Three types of ideal systems were constructed: clustered system, uniform system and hypothetical system. Comparing the values of the Hamming distance for all three types of ideal systems, we can conclude that the desired system best reflects the data analyzed for all integrated groups.

By comparing distribution of real universities with the desired system, we see that the percentage of universities placed, for example, in a group of elite universities, differs from the desired situation. This happens because we have divided the groups of universities based primarily on the average USE scores of the enrolled students, i.e. on the real existing data. Amongst the integrated groups, the number of such elite universities proves to be significantly lower than the model's prediction. This is where we need to grow, both for our universities and for our students.

Another delicate issue is the "removal" of higher education institutions. It should be noted that when we talk about "removing" universities from the system, we do not suggest closing these institutions, though work in this direction is being conducted at the state level. Among other things, the Monitoring aims to identify ineffective institutions. Moreover, one of the criteria of effectiveness is the average score of the students.[10] Not everything went smoothly with this monitoring in the first years, but this time the criteria of efficiency were clarified and their number was increased.[11]

---

[10] Interview with the Rector of the Higher School of Economics Yaroslav Kuz'minov. Kommersant, 2013, January 14. <http://www.kommersant.ru/doc/2101306>

In November 2013 the working groups of the Ministry of Education held several meetings where the results of the second monitoring of universities were discussed. As a result of these meetings about 70–80% of the branches of non-state and state universities may be suggested to close.[12] Yet there is another situation with head universities, placed to the list of inefficient institutions. If a university is the only one in a region and plays a crucial role in the regional educational system, it will be offered the chance to propose a development program. If these measures have no effect, the institution may be reorganized.[13]

However, we are concerned with the efficiency of universities only indirectly. We are interested instead in rather narrow questions: what is the heterogeneity that the Russian system of universities work with? What happens to the heterogeneity if, theoretically speaking, the weakest element would be "eliminated" from the system?

Obviously, the preferable situation is if universities would begin to do everything possible to attract a student body with a higher level of preparation. But this is not always a question that universities themselves can resolve. Here, numerous questions emerge regarding the quality of education, the influence of teachers, family, etc. These are the sort of questions that have no simple answers.

## 7. Conclusion

We have proposed a new method of studying heterogeneity in the higher education system. Our method is based on the three models of the hypothetical educational system, and their comparison to the real systems.

---

[11] The second stage of monitoring schools. Rossiyskaya Gazeta, 2013, November 18. <http://www.rg.ru/2013/11/18/monitoring-site.html>

[12] Ministry of Education wants to close 80 percent of the universities' branches. Rossiyskaya Gazeta, 2013, December 2. < http://www.rg.ru/2013/12/02/filial-site.html>

[13] Three Novosibirsk universities can be reorganized after monitoring of the Ministry of Education. RIA Novosti, 2013, November 25. < http://ria.ru/nsk/20131125/979547840.html>

We applied this method to study four groups of majors of Russian universities and obtained interesting results concerning the heterogeneity of the system of Russian universities. However, we are dealing with a fairly narrow problem. We try to build the assessment of higher education institutions on the basis of only one parameter: input grades of university students. The next questions follow naturally: how can we extend the basis for the evaluation of the heterogeneity of the entire system of higher education? Which are the additional parameters to be used in this evaluation?

# Appendix

## 1. A procedure of the missed values' replacement

The number of missing values in our database is relatively small. Basically missing values imply average scores of students enrolled in universities as winners of student Olympiads (i.e. without regard to their points of USE). Several universities did not provide information about the USE scores of the students admitted on the basis of targeted contracts with enterprises and regional authorities, or on the basis of benefits. Some universities did not post information about the USE scores of tuition-based students enrolled on some specialties.

Percentages of missed values in each of four aggregated groups of majors are shown in Table 1.

**Table 1.** The percentage of missing values

| Group of majors | Electronic Engineering | Economy and Management | Agriculture and Fishery | Healthcare |
|---|---|---|---|---|
| Total number of missing values, % | 3,2 | 2,5 | 1,59 | 0 |

In an ideal situation, the presence of even such a small number of missing values requires additional procedures for their processing. There is a number of options to substitute missing values: replacement averages, calculation of possible values using the regression models and several others. According to [Tabachnick, Fidell, 2001], if the percentage of missing values does not exceed 5%, they could be ignored.

So, in order to solve the problem of incomplete set of characteristics in our study we developed a special technique of the missing values replacement presented as follows.

There are data for $k$ students of $n$ universities. For each student we know
1) University.
2) The form of admission (state-funded or tuition-based).
3) The basis of admission (on competition, out of competition, etc.).

4) Unified State Examination (USE) mean score (is not known for all students).

There are gaps in USE scores of a certain number of students. Let us describe the procedure of filling these missing values.

Let us denote the set of all universities as $U=\{u_1, u_2, ..., u_n\}$, where $u_i$ is $i$-th university, and the set of all students $S=\{s_1, s_2, ..., s_k\}$, where $s_j$ is $j$-th student. Each student $s_j$ is assigned a vector $Z^j=(z_1^j, z_2^j, z_3^j, z_4^j)$, characterizing each student by four features described above. If we do not know USE score of $j$-th student, then $z_4^j = 0$.

Let the set of all USE points be $E=\{e^{i_1}, e^{i_2}, ..., e^{i_k}\}$, where the index $i$ corresponds to the $i$-th university (if the USE score is unknown, it is set to equal to 0). The set $E$ is further divided into $n$ subsets $E^i$ ($n$ is the total number of universities), where index $i$ correspond to $i$-th university. Each of the subsets $E^i$ is subdivided into two subsets: $E^i_{bud}$ – containing estimated exam score of students admitted to the state-funded form of training, and $E^i_{com}$ – containing estimated exam score of students admitted to a tuition-based form of funding. Zero values of the set $E^i$ are included into the set $E^i_{pr}$ (the set of score gaps of all students), zero values of set $E^i_{bud}$ are included into the set $E^i_{pr\_bud}$ (the set of score gaps of the state-funded students), and zero values of the set $E^i_{com}$ are included into the set $E^i_{pr\_com}$ (the set of score gaps of the tuition-based students).

For each university $i$ it is calculated the total number of students $k^i$ (where $i$ – the index of the $i$-th university), total number of state-funded students $k^i_{bud}$, total number of tuition-based students $k^i_{com}$, total number of students with gaps in USE scores $k^i_{pr}$, total number of state-funded students with gaps in USE scores $k^i_{pr\_bud}$, and total number of tuition-based students with gaps in USE scores $k^i_{pr\_com}$

$$k^i = |E^i|, \qquad (1)$$
$$k^i_{bud} = |E^i_{bud}|, \qquad (2)$$
$$k^i_{com} = |E^i_{com}|, \qquad (3)$$
$$k^i_{pr} = |E^i_{pr}|, \qquad (4)$$
$$k^i_{pr\_bud} = |E^i_{pr\_bud}|, \qquad (5)$$
$$k^i_{pr\_com} = |E^i_{pr\_com}|. \qquad (6)$$

Let us introduce some new notations: $U'$ – the set of the excluded universities, $S'$ – the set of the excluded students. For each university $u_i \in U$ the conditions (7) and (8) are tested

$$k^i < 15, \tag{7}$$

$$k^i_{pr} \geq 0{,}25 * k^i. \tag{8}$$

For Agriculture universities condition (7) is $k^i<8$. With the assumption that at least one of the conditions (7) or (8) holds, the university $i$ is included into the set $U'$. All students of the university $i$ are included into the set $S'$. The USE points $e^{ij}$ of the students $s_j \in S'$ are included into the set $E'$ (the index $j$ corresponds to the $j$-th student). The number of excluded universities $n_{excl}$ is calculated as $n_{excl} = |E'|$.

All the elements of the set $U'$ are excluded from the set of $U$, i.e. further we will consider the set $U^* = U \setminus U'$. The same procedure is carried out for sets $S$ and $E$, i.e. $S^* = S \setminus S'$, $E^* = E \setminus E'$. Similarly, the procedure is done for the subsets of $E$, $E^*$ the set is divided into $n - n_{excl}$ subsets $E^{i*}$, corresponding to USE points of state-funded students of university $u_i \in U^*$. Subsets of $E^{i*}$ are: $E^{i*}_{bud}$ – containing estimated exam score of students admitted to the state-funded form of training, and $E^{i*}_{com}$ – that of students admitted to a tuition-based form of funding. For exam scores of students of universities $u_i \in U^*$ we find the maximum and minimum values of the USE on the state-funded ($max^i_{bud}$ and $min^i_{bud}$) and tuition-based ($max^i_{com}$ and $min^i_{com}$) as follows

$$max^i_{bud} = max(E^{i*}_{bud}), \tag{9}$$

$$min^i_{bud} = min(E^{i*}_{bud}), \tag{10}$$

$$max^i_{com} = max(E^{i*}_{com}), \tag{11}$$

$$min^i_{com} = min(E^{i*}_{com}). \tag{12}$$

Given the number of state-funded and tuition-based students of each university, and their USE scores, we calculate average values of USE scores for each university for state-funded ($e^i_{sr\_bud}$) and tuition-based ($e^i_{sr\_com}$) forms of admission

$$e^i_{sr\_bud} = \frac{1}{k^i_{bud} - k^i_{pr\_bud}} \sum_{j=1}^{k^i_{bud}} e^{ij}, \qquad (13)$$

$$e^i_{sr\_com} = \frac{1}{k^i_{com} - k^i_{pr\_com}} \sum_{j=1}^{k^i_{com}} e^{ij}. \qquad (14)$$

For each university the 10% range from the maximum values of USE scores for the state-funded and for the tuition-based forms is constructed:

$$max^i_{bud+10} = max(E^i_{bud}) + max(E^i_{bud})*0{,}1, \qquad (15)$$

$$max^i_{com+10} = max(E^i_{com}) + max(E^i_{com})*0{,}1, \qquad (16)$$

$$min^i_{bud-10} = max(E^i_{bud}) - max(E^i_{bud})*0{,}1, \qquad (17)$$

$$min^i_{com-10} = max(E^i_{com}) - max(E^i_{com})*0{,}1. \qquad (18)$$

If $max^i_{bud+10} > 100$, then $max^i_{bud+10} = 100$. The same holds for $max^i_{com+10}$ (if $max^i_{com+10} > 100$, then $max^i_{com+10} = 100$).

Next, for each university $u_i \in U^*$ the variance $\delta^2$ is computed (separately for the state-funded and for the tuition-based)

$$\delta^2_{bud} = \frac{1}{k^i_{bud} - k^i_{pr\_bud}} \sum_{j=1}^{k^i_{bud}} (e^{ij} - e^i_{sr\_bud}), \qquad (19)$$

$$\delta^2_{com} = \frac{1}{k^i_{com} - k^i_{pr\_com}} \sum_{j=1}^{k^i_{com}} (e^{ij} - e^i_{sr\_com}). \qquad (20)$$

The intervals for filling the gaps of all students are calculated with the exception for those who entered universities by Olympiad competition, according to formulas (21) – (24). Formula (21) describes the lower limit of the state-funded students ($min^i_{bud\_dis}$), formula (22) – the upper limit of the state-funded students ($max^i_{bud\_dis}$), formula (23) – the lower limit of the tuition-based students ($min^i_{com\_dis}$), and (24) describes the upper limit of the tuition-based students ($max^i_{com\_dis}$)

$$min^i_{bud\_dis} = e^i_{sr\_bud} - \sqrt{\delta^2_{bud}}, \tag{21}$$

$$max^i_{bud\_dis} = e^i_{sr\_bud} + \sqrt{\delta^2_{bud}}, \tag{22}$$

$$min^i_{com\_dis} = e^i_{sr\_com} - \sqrt{\delta^2_{com}}, \tag{23}$$

$$max^i_{com\_dis} = e^i_{sr\_com} + \sqrt{\delta^2_{com}}. \tag{24}$$

After completing all calculations, the missing values are filled out. In case a student entered a university due to the results of the Olympiad competition, the USE scores are set randomly in the range [$min^i_{bud-10}$; $max^i_{bud+10}$]. For those tuition-based students who entered a university on a regular basis the USE scores are given randomly in the range ($min^i_{com\_dis}$; $max^i_{com\_dis}$). For state-funded students who entered a university on a regular basis the USE scores are given randomly in the range of ($min^i_{bud\_dis}$; $max^i_{bud\_dis}$).

## 2. Main characteristics of 4 integrated groups of majors

### *2.1. Universities within the integrated group of Majors "Electronic Engineering, Radio Engineering and Communications"*

Figure A1 shows the distribution of state-financed and tuition-based students in each university within the integrated group of majors "Electronic Engineering" (all universities are numbered from 1 to 105).

Minimum number of state-financed students is 15, maximum is 609. Minimum number of tuition-based students is 0, maximum is 85. The majority of universities have no more than 200 state-financed students, while the number of tuition-based students is quite small.

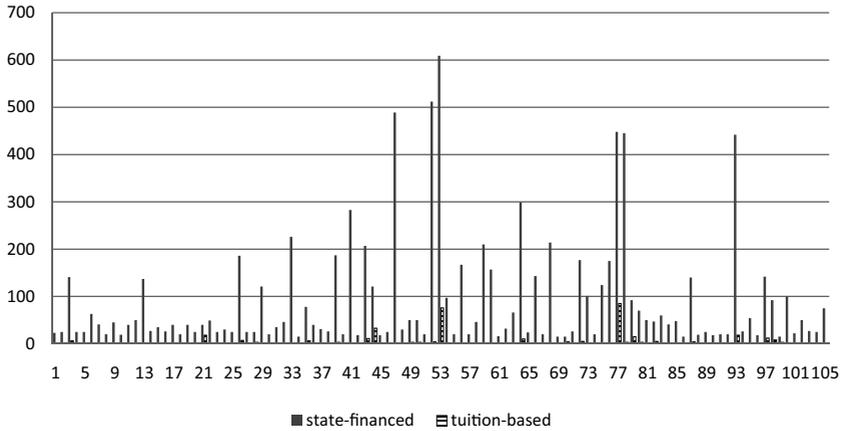

**Figure A1.** The distribution of state-financed and tuition-based students within the integrated group of majors "Electronic Engineering"

Figure A2 shows the range of maximum and minimum Unified State Examination (USE) scores of the state-financed students for each university within the integrated group of majors "Electronic Engineering". As it can be seen in the following Fig. 2, the range of scores is quite big.

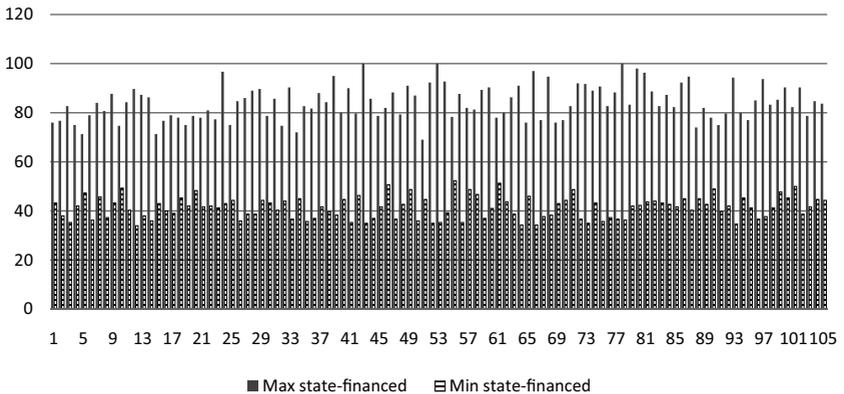

**Figure A2.** The range of maximum and minimum USE scores of state-financed students within the integrated group of majors "Electronic Engineering"

Figure A3 shows the average USE scores of state-financed students within the integrated group of majors "Electronic Engineering".

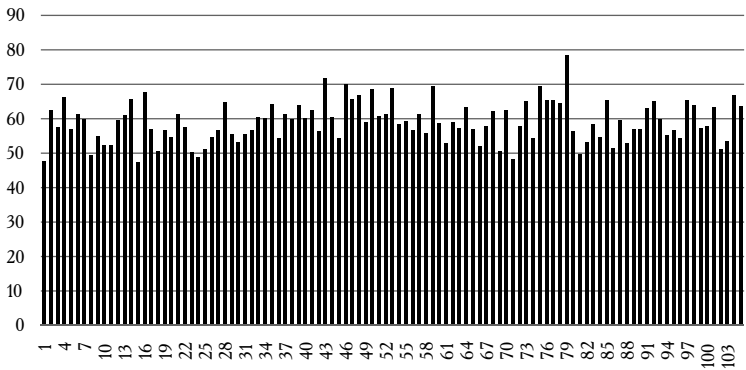

**Figure A3.** The average USE scores of the state-financed students within the integrated group of majors "Electronic Engineering"

Figure A4 shows the average USE scores of tuition-based students within the integrated group of majors "Electronic Engineering".

Value "0" means the absence of tuition-based students in the university in this area.

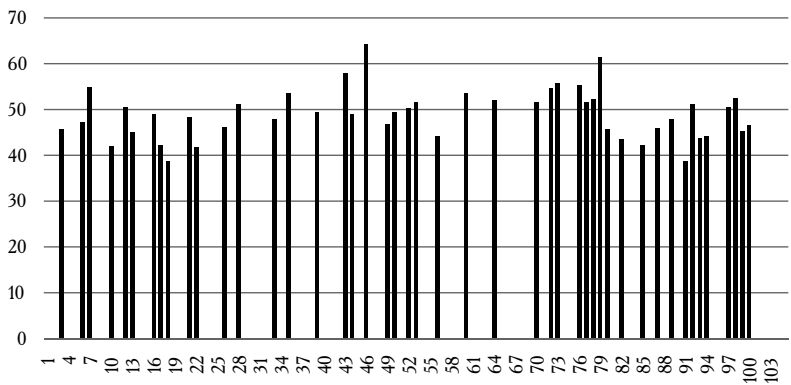

**Figure A4.** The average USE scores of tuition-based students within the integrated group of majors "Electronic Engineering"

As can be seen on Figures A3 and A4 the distribution of scores of state-financed and tuition-based students across universities are quite "flat".

## 2.2. Universities within the integrated group of Majors "Economics and Management"

Figure A5 shows the distribution of state-financed and tuition-based students in each university within the integrated group of majors "Economics and Management" (all universities are numbered from 1 to 378).

Minimum number of state-financed students is 0, maximum – 773. Minimum number of tuition-based students is 0, maximum – 1335. As can be seen from Fig. A5 the number of tuition-based students is very large within the studied group of majors.

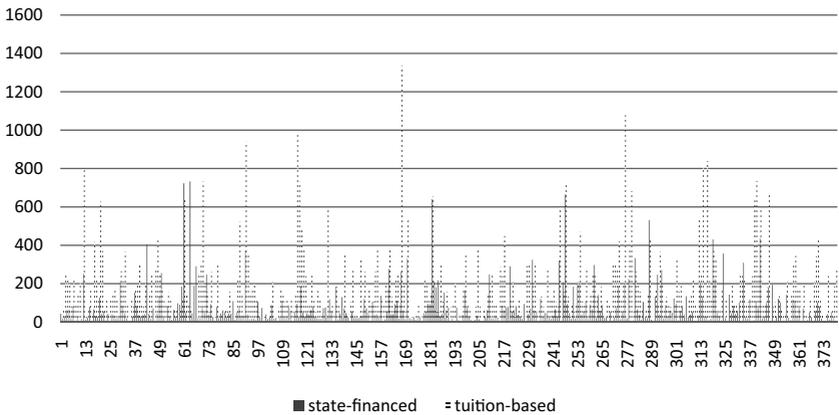

**Figure A5.** The distribution of state-financed and tuition-based students in universities within the integrated group of majors "Economics and Management"

Figure A6 shows the range of maximum and minimum Unified State Examination (USE) scores of the state-financed students for each university. The range in scores is observably big.

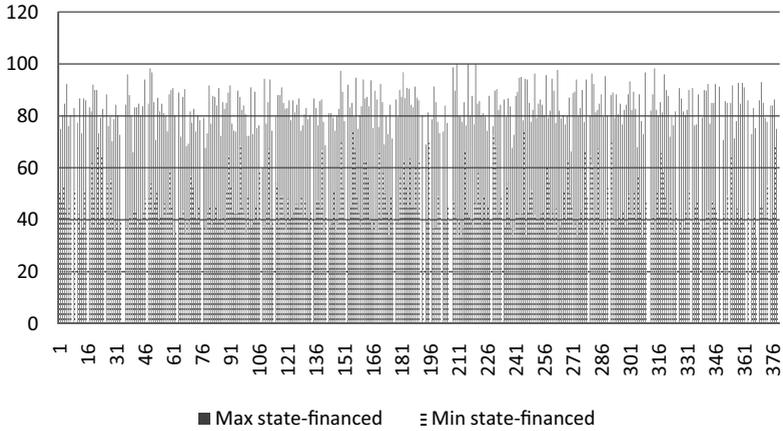

**Figure A6**. The range of maximum and minimum USE scores of state-financed students within the integrated group of majors "Economics and Management"

Figure A7 shows the average USE scores of state-financed students. Value "0" shows the absence of state-financed students in a university in this area.

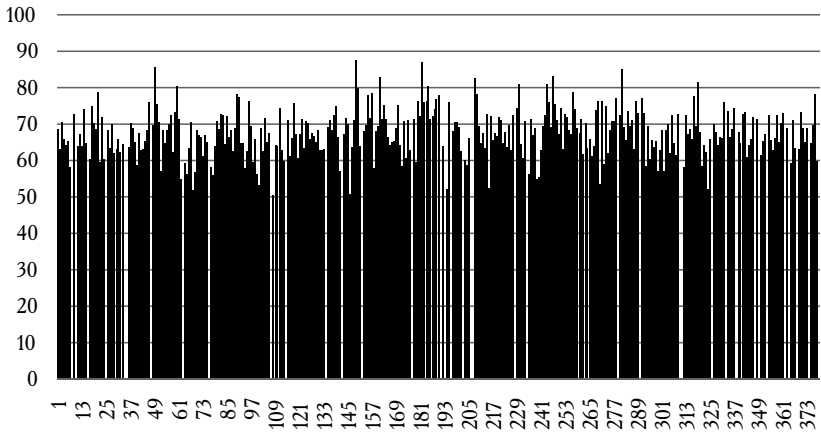

**Figure A7.** The average USE scores of state-financed students within the integrated group of majors "Economics and Management"

Figure A8 shows the average USE scores of tuition-based students. Value "0" means the absence of tuition-based students in a university in this area.

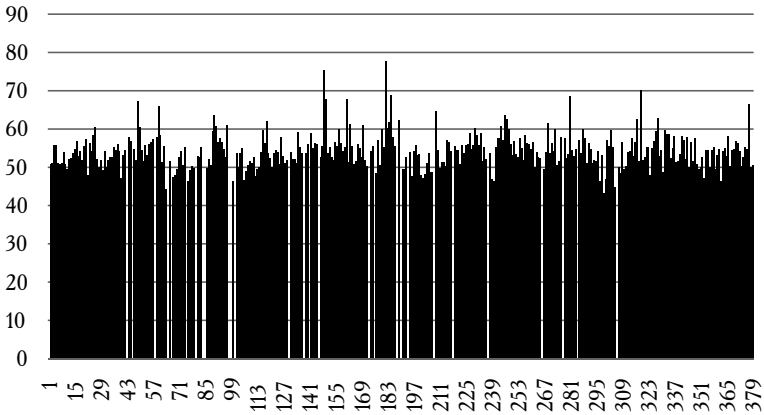

**Figure A8.** The average USE scores of tuition-based students within the integrated group of majors "Economics and Management"

Figures A7 and A8 show that the difference between scores of tuition-based students and state-financed students is quite big. Those universities vary markedly in the quality of students they admit.

## 2.3. Universities within the integrated group of Majors "Agriculture and Fishery"

Figure A9 shows the distribution of state-financed and tuition-based students in each university within the integrated group of Majors "Agriculture and Fishery" (all universities are numbered from 1 to 85).

The minimum number of state-financed students is 5, maximum – 561. Minimum number of tuition-based students is 0, maximum – 212.

As can be seen many universities do not have tuition-based students, or their number is very small.

Figure A10 shows the range of maximum and minimum Unified State Examination (USE) scores of students for each university within the integrated group of Majors "Agriculture and Fishery".

The range of scores looks very big, and the minimum score in the majority of universities is extremely small.

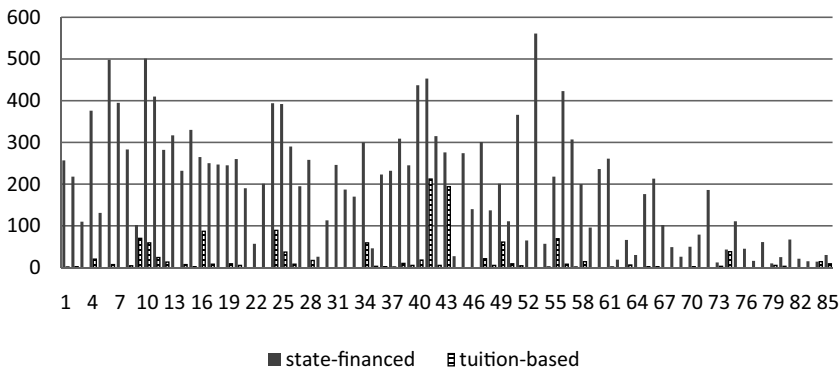

**Figure A9.** The distribution of state-financed and tuition-based students in universities within the integrated group of majors "Agriculture and Fishery"

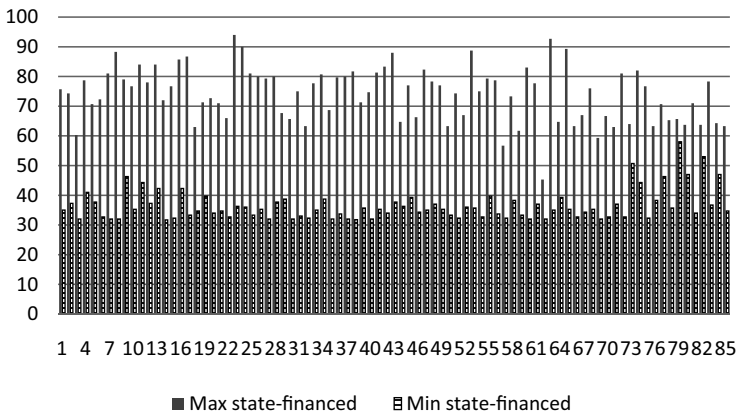

**Figure A10.** The range of maximum and minimum USE scores of state-financed students within the integrated group of majors "Agriculture and Fishery"

Figure A11 shows the average USE scores of state-financed students in each university. The scores look quite homogenous.

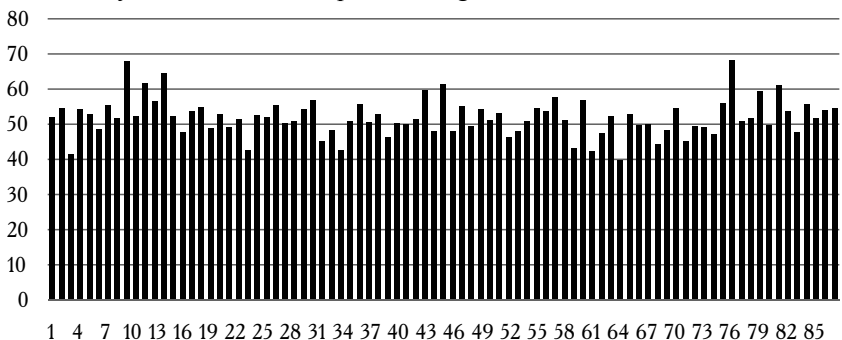

**Figure A11.** The average USE scores of state-financed students within the integrated group of majors "Agriculture and Fishery"

Figure A12 shows the average USE scores of tuition-based students. Value "0" means the absence of tuition-based students in the university in this area. Universities vary observably in quality of tuition-based students they take.

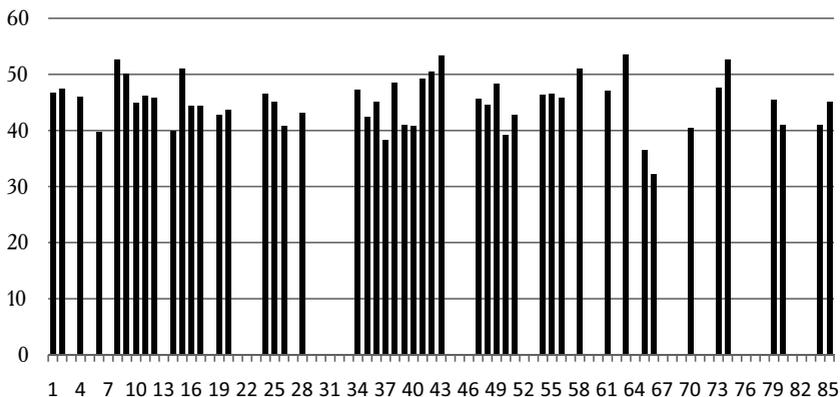

**Figure A12.** The average USE scores of tuition-based students within the integrated group of majors "Agriculture and Fishery"

## 2.4. Universities within the integrated group of Majors "Healthcare"

Figure A13 shows the distribution of state-financed and tuition-based students in each university within the group of Majors "Healthcare" (universities are numbered from 1 to 77).

The minimum number of state-financed students is 20, maximum – 1043. Minimum number of tuition-based students is 0, maximum – 1390.

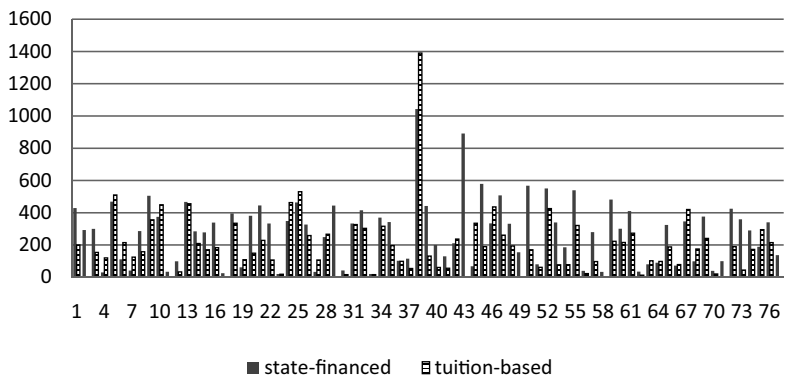

**Figure A13.** The distribution of state-financed and tuition-based students in universities within the integrated group of majors "Healthcare"

Figure A14 shows the range of maximum and minimum Unified State Examination (USE) scores of the students for each university.

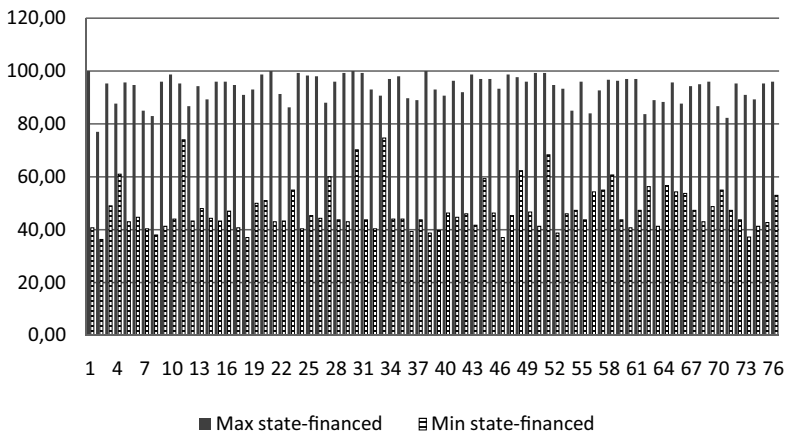

**Figure A14.** The range of maximum and minimum USE scores of state-financed students within the integrated group of majors "Healthcare"

Figure A15 shows the average USE scores of state-financed students in each university.

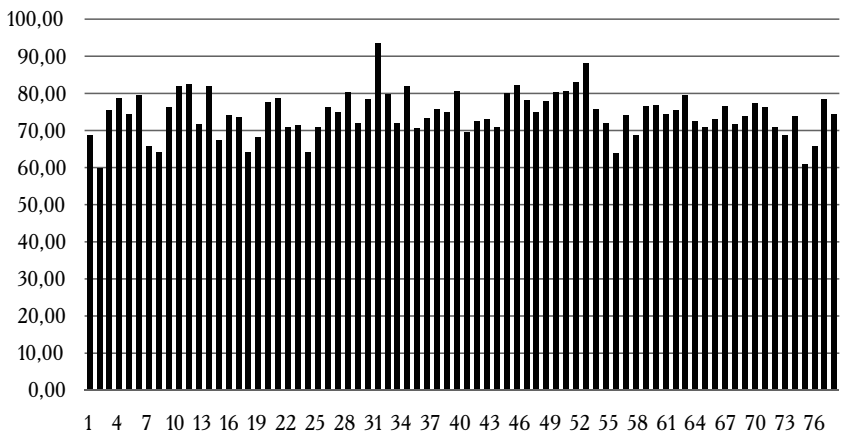

**Figure A15.** The average USE results of state-financed students within the integrated group of majors "Healthcare"

As can be seen from the Figure above the scores of state-financed students in medical specialties are in general higher than 70 and quite homogeneous.

Figure A16 shows the average USE scores of tuition-based students. Value "0" means the absence of tuition-based students in a university in this area.

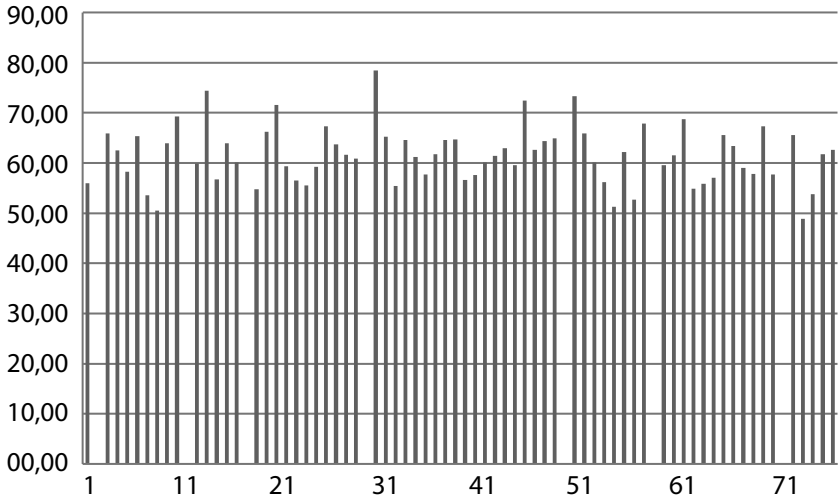

**Figure A16.** The average USE results of tuition-based students in universities within the integrated group of majors "Healthcare"

**Неоднородность образовательных систем: введение в проблему** [Текст]: препринт WP7/2014/03 / Ф. Т. Алескеров, И. Д. Фрумин, Е. Ю. Карданова и др. ; Нац. исслед. ун-т «Высшая школа экономики». – М. : Изд. дом. Высшей школы экономики, 2014. – (Серия WP7 «Математические методы анализа решений в экономике, бизнесе и политике»). – 64 с. – 35 экз.Рассматривается проблема неоднородности образовательной системы на базе единственного параметра – входных баллов студентов. Предлагается математическая модель системы университетов, основанная на построении интервального порядка. Для оценки неоднородности системы применяется расстояние Хемминга, и для иллюстрации того, как наша модель работает, мы используем баллы ЕГЭ российских студентов. Полученные результаты позволяют увидеть, что слабые вузы делают всю систему университетов гетерогенной и слабоструктурированной. И наоборот, после удаления слабой составляющей оставшийся пул вузов превращается в хорошо согласованную систему.
*Авторы*:
*Алескеров Фуад Тагиевич*, НИУ ВШЭ, Отделение прикладной математики и информатики, научный руководитель
*Фрумин Исак Давидович*, НИУ ВШЭ, Институт образования, научный руководитель
*Карданова Елена Юрьевна*, НИУ ВШЭ, Центр мониторинга качества образования, директор
*Иванова Алина Евгеньевна*, НИУ ВШЭ, Центр мониторинга качества образования, младший научный сотрудник
*Мячин Алексей Леонидович*, НИУ ВШЭ, Международная научно-учебная лаборатория анализа и выбора решений, научный сотрудник; Российская академия наук, Институт проблем управления, научный сотрудник
*Серова Александра Владимировна*, НИУ ВШЭ, Методический центр, руководитель центра
*Якуба Вячеслав Иванович*, Российская академия наук, Институт проблем управления, старший научный сотрудник

Ф.Т. Алескеров, И.Д. Фрумин, Е.Ю. Карданова, А.Е. Иванова,
А.Л. Мячин, А.В. Серова, В.И. Якуба

**Неоднородность образовательных систем:
введение в проблему**

(*на английском языке*)